\documentclass[conference,compsoc]{IEEEtran}

\ifCLASSOPTIONcompsoc
  \usepackage[nocompress]{cite}
\else
  \usepackage{cite}
\fi

\usepackage{graphicx}

\begin{document}
%
\title{Blockchain for the IoT: Opportunities and Challenges}

\author{\IEEEauthorblockN{Gowri Sankar Ramachandran}
\IEEEauthorblockA{University of Southern California\\
Email: gsramach@usc.edu}
\and
\IEEEauthorblockN{Bhaskar Krishnamachari}
\IEEEauthorblockA{University of Southern California\\
Email: bkrishna@usc.edu}
}

\maketitle

\begin{abstract}
Blockchain technology has been transforming the financial industry and has created a new crypto-economy in the last decade. The foundational concepts such as decentralized trust and distributed ledger are promising for distributed, and large-scale Internet of Things (IoT) applications. However, the applications of Blockchain beyond cryptocurrencies in this domain are few and far between because of the lack of understanding and inherent architectural challenges. In this paper, we describe the opportunities for applications of blockchain for the IoT and examine the challenges involved in architecting Blockchain-based IoT applications.
\end{abstract}

\IEEEpeerreviewmaketitle

\section{Introduction}
Satoshi Nakamoto laid the foundation for the Blockchain technology in 2008 by presenting a solution for decentralized trust among unknown entities~\cite{nakamoto2008bitcoin}. BitCoin, the first decentralized digital currency, impacted financial institutions, and a wide-number of cryptocurrencies entered the market in the following years. The majority of blockchain applications currently involve digital cryptocurrencies, where the users exchange monetary value with each other through the decentralized framework. 

Enabling decentralized trust through a consensus protocol and distributed storage through a tamper-proof ledger are the critical features of the Blockchain technology. Any application that involves multiple stakeholders can benefit from these features because it enables transparent interactions without requiring a trusted third party. IoT applications in the context of smart cities and supply chain management consist of numerous stakeholders, where the Blockchain technology can be used to strengthen the confidence among the involved entities and organizations. 

Although the technology has been around for almost a decade, its technical underpinnings are made clearer only in the last two years. On the one hand, architects designing IoT applications are fully aware of the limitations and capabilities of contemporary IoT platforms and technologies. On the other hand, Blockchain developers and enthusiasts understand the practical details of the Blockchain frameworks and their viability on different classes of computation and storage platforms. We notice a gap between the two communities, and it is essential to bridge this gap to fully exploit the capabilities of blockchain technology beyond cryptocurrencies and FinTech applications.

This paper presents the promises of Blockchain for IoT and describes the challenges and limitations of the blockchain by correlating the architectural elements of IoT with the Blockchain. Furthermore, the paper also discusses the fundamental design questions for the application developers who are designing and implementing applications at the intersection of Blockchain and IoT.

Section~\ref{sec:iot} provides an overview and the architecture of IoT. Building blocks and architectural elements of the blockchain are presented in Section~\ref{sec:blockchain}. Section~\ref{sec:opp} discusses the opportunities for applying blockchain for the IoT. Section~\ref{sec:challenges} describes the challenges and open questions. Finally, Section~\ref{sec:con} concludes the paper.
\section{Overview of the IoT}
\label{sec:iot}
\begin{figure*}[t]
\centering
\includegraphics[scale=0.5]{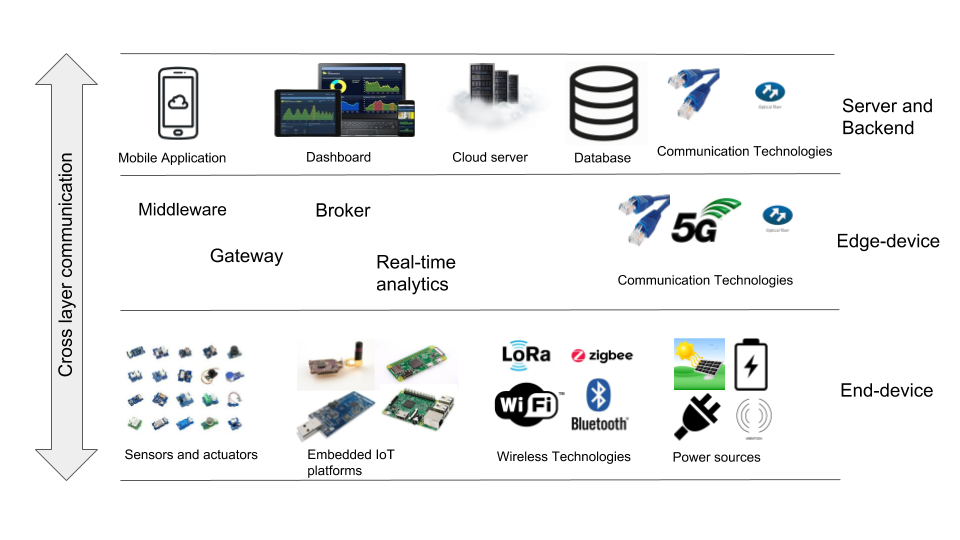}
\centering
\caption{Layered architecture of the Internet-of-Things applications.}
\label{fig:arch}
\end{figure*}

Application areas of IoT include air quality monitoring, smart cities, supply chain management, and production line monitoring. Internet-of-Things comprises of computation, communication, sensing, and actuation functionalities, and such functionalities are distributed throughout the network. IoT architecture can be broadly classified into three layers as shown in Figure~\ref{fig:arch}. 

\textbf{End-device Layer:} The end-device layer comprises of sensors, low-power embedded platforms, wireless communication technologies, and power sources. Low-power embedded IoT platforms act as a hub for sensors and one or more wireless communication technologies. IoT platforms are typically deployed in challenging and hard-to-reach environments. Thus, it is essential to keep the devices running longer on battery power or harvested energy. IETF defines the devices in this layer as very constrained sensor motes with limited processing and storage capabilities, and they are referred to as \emph{class 0} devices. The end-device layer is the most resource-constrained layer in IoT application architecture.

\textbf{Edge-device Layer:} The edge-device layer is responsible for collecting sensor data from end-devices. This layer consists of a network gateway for handling inbound and outbound communications with the end-device layer. Also, the data from multiple end-devices are processed in this layer to meet the real-time demands of the application. Devices at this layer are more capable than the end-device layer with respect to computation and storage capabilities.

\textbf{Server or backend layer:} The server or the cloud backend layer is responsible for storage and visualization functionalities. End-users of the IoT application interact with this layer for monitoring and control of their infrastructure. Web servers, data analytics engines, and databases operate at this layer to cater the demands of the end-users. Devices in this layer have the maximum processing and storage capacities among all the layers in the stack.

Table~\ref{tab:resource} summarizes the resource capacities at different layers of the IoT stack. The end-device layer is a constrained layer with insufficient resources for computation, communication, and storage, while the server or the backend layer consists of maximum resources. 

\emph{Application of any new technology and protocol to the IoT must consider the resource capacities of different layers before their deployment.}

IoT applications following the above architecture have been widely used in various deployments, but the integration of blockchain into such an architecture remains challenging as discussed in Section~\ref{sec:challenges}. The overview of blockchain and its fundamental building blocks are presented in the next section.

\section{Overview of the Blockchain technology}
\label{sec:blockchain}
\begin{figure*}[t]
\centering
\includegraphics[scale=0.4]{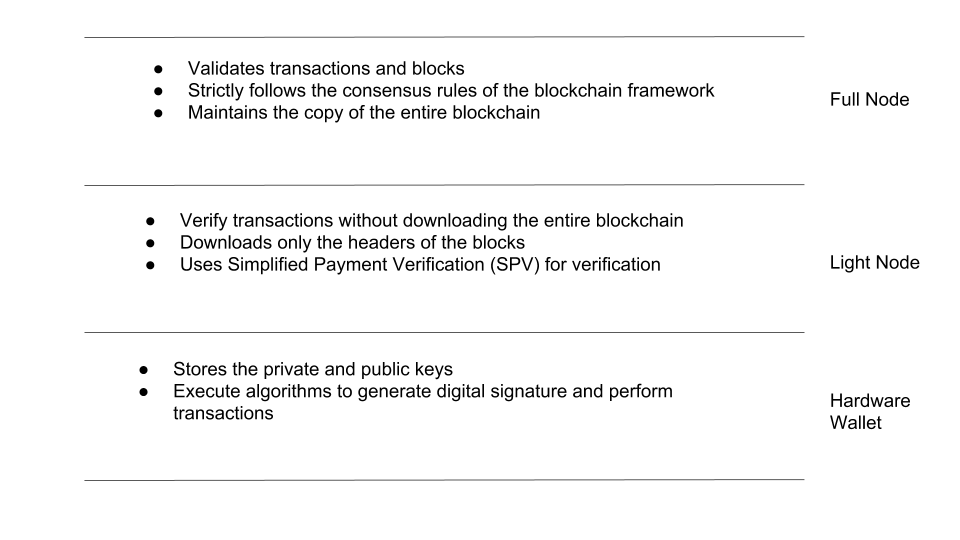}
\centering
\caption{Components of Blockchain.}
\label{fig:comp}
\end{figure*}

\begin{table*}[]
\centering
\caption{Resource requirements of IoT and Blockchain.}
\label{tab:resource}
\begin{tabular}{ccccc}
\hline
\textbf{Layers}    & \textbf{CPU} & \textbf{Memory} & \textbf{Power Budget} & \textbf{Bandwidth} \\ \hline
\multicolumn{5}{c}{Internet-of-Things (IoT)}                                                     \\ \hline
End-device         & Low          & Low             & Low                   & Low                \\ \hline
Edge-device        & Medium       & Medium          & High                  & High               \\ \hline
Server and Backend & High         & High            & High                  & High               \\ \hline
\multicolumn{5}{c}{Blockchain}                                                                   \\ \hline
Hardware Wallet    & Low          & Low             & Low                   & Low                \\ \hline
Light Node         & Medium       & Medium          & High                  & High               \\ \hline
Full Node          & High         & High            & High                  & High               \\ \hline
\end{tabular}
\end{table*}

The Blockchain technology uses the combination of cryptography, a consensus algorithm, and a distributed ledger to create a decentralized and trustworthy platform. In this section, we will discuss the three key aspects of Blockchain technologies.

\textbf{Cryptographic Digital Signature}: The public-key cryptography is used in blockchain to generate a signature for Blockchain transactions. Users carry out transactions by creating a digital signature using their private keys. Recipients in the blockchain network verify the transaction using the public key of the sender to ensure that the transaction is indeed signed by the sender. Source or end-devices sign the transactions when they create a transaction. 

\textbf{Distributed Ledger}: Blockchain use a distributed storage to record the transactions. In essence, all the platforms in the network store either the entire transactions or a subset of transactions. All the nodes in the network come to a consensus (using a consensus algorithm) before entering the transactions into the ledger. This feature makes blockchain effectively immutable.

\textbf{Consensus algorithm}: Blockchain does not rely on a centralized server for verification and validation of transactions. Instead, Blockchain uses a peer-to-peer model, and all the decisions within the network are made by the participating members through a consensus protocol. 

Figure~\ref{fig:comp} shows the components of Blockchain. The core functionalities of Blockchain are distributed across the hardware wallet, light nodes, and full nodes in a blockchain network. Table~\ref{tab:resource} presents the resource requirements of blockchain at different layers.

\section{Opportunities}
\label{sec:opp}

The core building blocks of blockchain such as public-key cryptography, distributed ledger, and consensus algorithms are promising for the IoT. We will describe the opportunities for applying blockchain technologies to the IoT in this section.

\begin{itemize}
\item \textbf{Privacy/anonymity:} Transactions in Blockchain use the digital identity generated using public-key cryptography and a hashing algorithm. IoT applications with sensitive information can leverage these mechanism to hide real identity in the network.
\item \textbf{Monetary exchange of data and compute:} Monetary exchange of data and compute: IoT applications in the area of smart cities use sensors in combination with crowdsourcing to deliver digital services to the city population. Monetary rewards may be essential to involve the community members in smart city applications and to leverage the edge resources such as computation power, storage, and bandwidth. Blockchain technology can also be used to set up a monetary system to issue tokens to the community members for their participation.
\item \textbf{Record transactions for account and audit:} The data from IoT applications are transported through infrastructure owned by multiple organizations. Supply chain monitoring focuses on tracking and monitoring assets throughout the supply chain process. Traditional supply chain monitoring systems rely on a centralized architecture, wherein all the data from assets are stored in a central database. Using blockchain for recording the data in a decentralized ledger increases the trust while moving assets (real or digital) through infrastructure owned by multiple and diverse stakeholders.
\item \textbf{Smart Contracts} Nick Szabo introduced the concept of Smart Contracts~\cite{szabo1994smart} as an alternative to the traditional paper-based contracts. A smart contract is a digital contract embedded in the system, which gets executed when the conditions declared in the agreement are met. Smart contracts arbitrate transactions autonomously while exchanging assets between parties or dealing with non-trusted members in a blockchain network. IoT applications, for example, can employ smart contracts when transporting sensor data through infrastructures owned by multiple stakeholders and selling data produced by the sensors

\end{itemize}
\section{Challenges}
\label{sec:challenges}
We now discuss the challenges that arise in applying blockchain for the IoT.

\begin{itemize}
\item \textbf{Resource constraints:} IoT platforms have limited resources for computation, communication, and storage, while Blockchain technologies demand excessive resources. Class A low-power IoT platforms have less than 10 KB of data memory and less than 100 KB of program memory~\cite{RFC7228}, while a Blockchain node requires memory in the order of GBs~\cite{resource}. In addition, the computation requirements of consensus algorithms such as Proof-of-Work are well-beyond the capabilities of low-power, resource-constrained IoT devices. Contemporary Blockchain technologies are therefore ill-suited for such low-power IoT devices because their resource demands. From Figure~\ref{fig:arch}, the end-devices and edge-devices does not have the capacity to execute the Blockchain processes, and the server layer is ideally suited for contemporary Blockchain technologies. Such an approach might connect a centralized IoT deployment to a decentralized blockchain network, where the server layer of the IoT deployment acts as the entry point to the blockchain network. 
\item \textbf{Bandwidth requirements:} Platforms in the Blockchain network have to interact with other platforms in the network to participate in the consensus process. Due to the decentralized nature of the consensus process, platforms in the network exchange information about the blockchain to validate transaction and to create new blocks. IoT devices operating at end-device layer have severe bandwidth constraints, which also means the contemporary blockchain solutions are not suited for end-devices. Edge-devices and servers may have sufficient bandwidth, but it is important to note that the bandwidth requirement of blockchain may exceed the bandwidth requirement of the application itself, at least with most contemporary blockchain protocols.
\item \textbf{Security:} Blockchain technology follows a decentralized architecture, wherein all the devices in the network coordinate and cooperate with each other through pre-defined protocols. Thus, the devices stay connected to the blockchain network for participating in the consensus process. This always-connected feature makes IoT devices potentially more susceptible to security attacks. 
\item \textbf{Latency demands:} IoT applications typically consist of a collection of data producers and data consumers, and in some cases, the data consumers react to an event and perform an actuation. The introduction of Blockchain technology in this context may reduce the responsiveness if the data consumer may be required to wait for the conclusion of the consensus process before reacting to an event. Contemporary Blockchain technologies are not suitable for time-sensitive IoT applications that need fully confirmed transactions. 
\item \textbf{Transaction fees:} Most open Blockchain technologies charge a fee for transactions, and use it to rewarding the nodes involved in consensus process. IoT devices cannot store all the data to such a blockchain since storing the data to a blockchain incurs a transaction fee. If one wishes to put data from IoT devices on such a blockchain, it may need to be aggregated to reduce the transaction fees, but in this case it would be important to make sure that the aggregation process does not filter out essential information. Alternatively, an architecture where the data itself is transported off the chain and only hashed values or key transaction records are stored on the blockchain for verification and provenance purposes may be preferred. 
\item \textbf{Permissioned vs public:} Contemporary blockchain technologies can be broadly classified into two categories as public and permissioned blockchains. Public blockchains such as Bitcoin and Ethereum allow anyone to become a part of the network without any authorization. Anyone wishing to participate in the public blockchain can simply download and install the necessary frameworks, and this type of blockchain technologies require substantial resources for consensus process. Permissioned blockchains, on the other hand, consists of authorized members in the network. This type of blockchain may be suitable for IoT applications involving multiple known organizations as the network consists of authorized members, which open up opportunities for fast, higher-throughput consensus protocols.
\item \textbf{Partition tolerance for intermittently connected devices:} IoT applications in the space of supply chain monitoring consist of mobile devices with intermittent connectivity. Also, the end-devices running on batteries use duty cycling to prolong the lifetime. Furthermore, the devices operating on the wireless bands regulated by ETSI and FTC has to adhere to the bandwidth limitations enforced federal authorities. In such scenarios, the devices connect to a server or edge-device intermittently to exchange data. Assuming an architecture in which the server is acting as a lightweight node for recording the IoT data to a blockchain, the server has to download and store the headers of the blockchain to keep itself synchronized. For intermittently connected IoT devices, the cost of running a lightweight node for recording IoT data in a blockchain network may outweigh the benefits because of the bandwidth, computation, and storage costs. New blockchain protocols and frameworks are essential for reducing the infrastructure cost when using blockchain for recording IoT transactions and DAG-based protocols such as IOTA provide partition tolerance by making it easy to merge transactions from partitioned parts of the network. 
\item \textbf{Transaction Volumes} - these are quite severely limited on most current open, permissionless blockchain technologies, also preventing high volume sensor data applications from being carried directly on the blockchain.
\item \textbf{Physical interface weakness} - As cyber-physical systems, individual sensors and actuators could be hacked or misused to report false or erroneous data that gets logged on to the blockchain in an immutable fashion. 
\end{itemize}

\section{Related Work}
\label{sec:relatedwork}

Literature combining blockchain and IoT contributes security and privacy solutions. Kshetri~\cite{8012302} validates the applications of blockchain for securing the IoT. Tomer \emph{et al.} contribute CIoTA~\cite{2018arXiv180303807G} to detect anomalies in IoT applications. CIoTA applies the concepts of blockchain in combination with extensible Markov Model (EMM) to identify malicious activities. Dorri \emph{et al.}~\cite{2017arXiv171202969D} presents the gaps in contemporary security and privacy methods, and contribute LSB, a lightweight and scalable blockchain for IoT security and privacy. LSB's lightweight protocols reduce the bandwidth and computation costs. Pietro \emph{et al.}~\cite{2017arXiv171100540D} investigate the communication overhead of blockchain synchronization protocols for the IoT and highlight the uplink and downlink bandwidth demands. PlaTIBART~\cite{2017arXiv170909612W} is a testing framework to manage and deploy blockchain networks for transactive IoT applications. Hossein \emph{et al.}~\cite{2017arXiv170508230S} introduce a distributed data storage framework for IoT applications using the blockchain. ~\cite{2017arXiv170508230S} ensures that the IoT data ownership stays with the stakeholders. These papers address some of the challenges in described in Section~\ref{sec:challenges}, however, the architectural details and performance implications are not clearly addressed, especially for resource-constrained IoT platforms. 

The opportunities and challenges of applying blockchain for the IoT are presented in the literature. Huckle \emph{et al.}~\cite{HUCKLE2016461} discuss the applications of blockchain for monetizing IoT applications, but their work does not describe the challenges. Seyoung \emph{et al.}~\cite{7890132} demonstrates how blockchain can be used for storing sensor data using smart contracts. Canoscenti \emph{et al.}~\cite{7945805} reviews the use cases of the blockchain and highlights the open problems in integrity, anonymity, and adaptability when storing IoT data in a decentralized network. The authors of~\cite{2017arXiv171100540D} analyze the communication overhead of blockchain synchronization for the IoT. Unlike~\cite{7890132} and ~\cite{7945805}, our work focuses on architectural challenges and performance implications when using blockchain for IoT for data storage, monetization, security, and privacy.

\section{Conclusion}
\label{sec:con}
Blockchain technology has already made a significant impact in the cryptocurrency applications. The fundamental building blocks - distributed ledger, consensus mechanisms, and public-key cryptography - of blockchain technology is promising for IoT and supply chain monitoring applications. We have discussed the architecture of IoT applications and mapped the functional blocks of the blockchain technology to reveal the architectural challenges involved in applying blockchain for the IoT. Next, we have presented opportunities for applying blockchain for the IoT. Finally, we concluded with the challenges which need to be addressed to fully exploit the benefits of blockchain technologies in the IoT domain. Despite the challenges, blockchain technologies are highly promising for resolving security, privacy, and trust issues in multi-stakeholder application environments.

\ifCLASSOPTIONcompsoc
  \section*{Acknowledgments}
\else
  \section*{Acknowledgment}
\fi
This work is supported by the USC Viterbi Center for Cyber-Physical Systems and the Internet of Things (CCI).

\bibliographystyle{IEEEtran}
\bibliography{IEEEabrv,references}

\end{document}